\documentclass[10pt, a4paper]{article}
 \usepackage{graphicx}                    
 \usepackage{color}                       
 \usepackage{url}                         
 
\begin{document}



Title: Relations between variability of the photosperic and interplanetary magnetic fields, solar wind and geomagnetic characteristics.

Authors: E. A. Gavryuseva (Institute for Nuclear Research RAS)

Comments:28 pages, 15 Postscript figures

\begin{abstract}
 Large scale solar magnetic field topology has a great influence 
 on the structure of the corona, heliosphera and geomagnetic perturbations.
 Data obtained over the last three solar cycles have been analysed
 to reveal the relationships between the photospheric field measured along the line of sight
 by the WSO group at 30 levels of heliolatitudes from -75 to 75 degrees and the interplanetary magnetic field

 The main aim of this first paper is to make a direct comparison between the basic structure and dynamics
 of the photospheric magnetic field and components and intensity of the interplanetary magnetic field 
 without using theoretical assumptions, models,
 physical expectations, etc.
 The second paper by Gavryuseva, 2018d presents the raports between different characteristics
 of the solar wind at the Earth orbit, and geomagnetic parameters
 provided by the OMNI team.

 The heliospheric and geomagnetic data are found to be
 divided into two groups characterized by their response to variability of the
 solar magnetic field latitudinal structures on short and on long time scales.
\end{abstract}
\vspace{1pc}
\section{Introduction}
   Large scale magnetic field of the Sun have an important role in driving
  perturbations of solar, interplanetary and geomagnetic conditions.

  Extensive investigations of the solar drivers
  was stimulated by the long-term rise of the geomagnetic
  activity over last century. Different explanations
  have been suggested, such as, for example, an increase of the
  strength of the interplanetary magnetic field 
  and solar wind speed and concentration   (Stamper et al.,  1999).
  This trend is seen in the photospheric field at low latitudes,
  as follows from Mt. Wilson Observatory magnetic data since 1967  (Li et al.,  2001).
  Makarov et al. (2002) found an increase of the polar cap area,
  which could contribute to the increase of the geomagnetic activity.
  Subsequently it was found that the contribution of the
  coronal holes is changing during the solar activity cycle
    (Luhmann et al.,  2002).
  It was established
  that around the minima (maxima) of solar activity the
  main drivers are polar (low latitude) magnetic field
   (Gonzalez et al.,  1994, 1999;  Wang et al.,  2000;   Wang et al.,  2000)
  but a deeper analysis is advisable taking into
  account the latitudinal structure of the large scale
  photospheric magnetic field (SMF).

     In previous papers  (Gavryuseva,  2006, 2006a, Gavryuseva, E., \& G. Godoli (2006),
     the relevance of the basic topology and dynamics
     of the photospheric magnetic field for the research of the solar drivers of the variability
     of solar wind and geomagnetic parameters has been underlined.

  Direct comparison between the mean latitudinal
  photospheric field variability, solar wind
  and geomagnetic perturbations
  on long and  short time scales was performed
  to search for the relations between them.
  Such a direct comparison of different parameters
  with solar magnetic field as a possible source of
  helio- and magneto-spheric perturbations is physically
  meaningful as well as practically useful,
  revealing the relations between them
  without introduction of theoretical  models,
  calculation of bilateral correlations,
  reliance on physical expectations, etc.

  This approach led to the conclusion that all solar wind data
  and geomagnetic perturbations that were examined divided into two groups
  characterized by sensitivity to the variability of the interplanetary
  magnetic field  and photospheric field at different latitudes.

   The data used are described in section 2: solar photospheric
  magnetic field (SMF) measurements from the Wilcox Solar Observatory (WSO) (section 2.1)
  and interplanetary magnetic  field ($IMF$), solar wind
  and geomagnetic parameters taken from
  the Operating Missions as Nodes on the Internet (OMNI) data base of
  NASA's Goddard Space Flight Center in Greenbelt (USA) (section 2.2).
  Very brief descriptions of the rotational variability of the $IMF$ 
  and short term variability of the interplanetary parameters are
  presented in sections 3 and 4.
  The delay between the $IMF$ and $SMF$ measurements at
  different latitudes is discussed in section 5.
  The cross-correlation between the $IMF$ and $SMF$ temporal behaviour within
  2-year long subsets of data running through the last three solar cycles is presented in section 6.
   Section 7 is devoted to the analysis of the relationships between the global structures
   of solar magnetic field, interplanetary field, solar wind parameters and
   geomagnetic perturbations on time scales from  1 year to 25 years 
   for measured data (section 7.1).
   The summary is presented in section 8.
 \section{Data}
 \subsection{Photospheric Magnetic Field}
  We analyse the temporal disturbances
  of the interplanetary magnetic field, 
  and their relationships with the  photospheric field of the Sun
  during the last three cycles of activity from 1976 to 2004.
  Daily data were used to find the optimum delay
  between the different parameters.

 We use  Wilcox Solar Observatory  data
 for the photospheric magnetic fields
 http://wso.stanford.edu/synopticl.html,
  (Scherrer et al.,  1977),
 OMNI data for solar wind parameters at the
 Earth's orbit, and indices of geomagnetic activity
 for the period 1976-2004,
 to study the relations between the solar wind,
 geomagnetic disturbances and
 solar drivers at different solar latitudes.

  In order to understand from which latitudinal zone
  the solar wind is originated and
  how it depends on the activity cycle
  it is necessary to know the latitudinal $SMF$ 
  structure over at least 22 years.

 The latitudinal structure of the $SMF$ has been deduced
 for the last 29 years since May 27, 1976
 from the Wilcox Solar Observatory (WSO) data
  (Scherrer et al.,  1977;
   Gavryuseva \& Kroussanova,  2003, Gavryuseva \& Gogoli, 2006, Gavryuseva, 2005, 2006, 2006a,b, 2008a,b, 2010 and references there).
 The structure in latitude and time of the 1-year running mean
 of the solar magnetic field
 with 1 Bartels Rotation (BR, 1 BR = 27 days) step is shown 
 on the upper plot in Fig. 1.

   The latitudinal four zonal ($4-zonal$) structure with a 22-year period
   with boundaries  of the polarity zones located at 25 S, 0,
   and 25 N degrees is clearly noticeable  (Gavryuseva, 2005, 2006, 2006a, 2008a, 2010).
   Yearly variability due to the Earth orbital rotation were removed 
   by the 1-year smoothing.

  The short-term variability of the solar filtered magnetic field ($FMF$)
  has  been  calculated as a residual between
  the $SMF$ smoothed by one and by two years.
 The short-term variability of the $SMF$ is shown on the bottom plot.
 The waves of the magnetic field  running through latitudes and 
 changing the polarity with a 2--3-year period ($WRL-$topology)
 are well visible on the bottom plot of Fig. 1
 (for more details see  Gavryuseva,  2005, 2006, 2006a, 2008a, 2010).
 In these plots yellow and red (blue) colors
 indicate positive (negative) polarity.
 The contours correspond to the levels of
 $-100, -50, 0, 50, 100$  $\mu$T.

   The antisymmetric field, equal to the difference of the magnetic field
   at the same latitudes in both hemispheres,
   clearly reveals the  magnetic field
  of opposite polarity of  the same magnitude
   at  certain latitudes.

   Symmetric $SMF$ structure corresponds to the mean level
   of the magnetic field of the same polarity
   at the latitudes $\theta$ and $-\theta$ in
   the northern and southern hemispheres.

 Daily mean data were used for
 day-to-day comparison with  solar wind and geomagnetic parameters.
 Fig. 2 and Fig. 3 show 
  the original $SMF$ distribution on the upper plot, the antisymmetric part on 
  and  the symmetric part of the solar magnetic field on the bottom plot in
  1996 (before the polarity inversion in the poles in 2000) and in 2003
  (after the polarity inversion during the  maximum of the solar activity in 
  The $SMF$ in 1996 was unusual, i.e. the positive polarity
  spread from the northern pole
  to about 45-50 degrees in the southern hemisphere
  during the  minimum of the solar activity.

  The 27-day variability of the  $SMF$  seen in Fig. 2 and Fig. 3
  is due to the solar rotation and
  its latitudinal dependence
  (the polar zones are rotating slower than the low latitude regions).
  The rotational periodicity  of the $SMF$ is more evident
  when the solar activity is high as 2003.
 \subsection{Interplanetary Magnetic Field, Solar Wind and Geomagnetic Data}
  The solar wind and geomagnetic  data were taken from the OMNI directory
  (http://nssdc.gsfc.nasa.gov/omniweb)
  which contains the Bartels mean values of the interplanetary magnetic  field 
 (IMF)
  and solar wind plasma parameters measured by various
  space-crafts near  the  Earth's  orbit,  as  well  as  geomagnetic
  and solar   activity indices).
  First, daily averages are deduced from OMNI's basic hourly values,
  and then the 27-day Bartels averages are deduced from 
  the daily averages.
  The corresponding standard deviations are related to only these averages
  and do not include the variances in the higher resolution data.


The $IMF$ and solar wind parameters taken into account
 are the following:\\
  $B_x$,   $B_y$,  $B_z$ and
  $B = (B_x^2+B_y^2+B_z^2)^{1/2}$
  are the components and magnitude
  of the interplanetary magnetic field, in nT;\\
  Proton density, $N_p$,   in $N/cm^3$;\\
  Proton temperature, $T_p$, in degrees  $K$;\\
  Plasma speed, $V_p$, in $km/s$;\\
  Electric field, in mV/m;\\
  Plasma beta,
   $N_{\beta}= [(T*4.16/10^5) + 5.34] * N_p/B^2$;\\
  Ratio   $N_{\alpha}/N_{p}$;\\
   Flow Pressure, $P$ proportional to $N_p*V^2$, in nPa;\\
   Alfven Mach number,      $M_a = (V*N_p^{0.5})/20*B$.\\
  The geomagnetic parameters taken into account are the following:\\
  $AE$-index;\\
  Planetary Geomagnetic Activity Index, $K_p-$ index;\\
  $DST$-index, in nT.\\
  Sunspot number ($SSN$) was used, as well, for a further comparison.

The $X$ axis directed along the intersection line of the ecliptic and solar equatorial 
planes to the Sun, the Z axis is directed perpendicular 
and north-ward from the solar equator, 
and the $Y$ axis completes the right-handed set. 

  The solar wind parameters
  analysed cover the same period as the WSO solar data
  with one  Bartels rotation  resolution.
  We call the set of these 16 parameters taken
  from the OMNI data base as "solar wind"
  ($SW$) data; they include the
  interplanetary magnetic field,
  solar wind and geomagnetic parameters and 
  sun spot number ($SSN$).
  These data are plotted in Fig. 4,                                                   
they are normalized to the maximal value and then
 smoothed by 13 Bartels rotations (about 1 year).
 \section{Rotational variability of Interplanetary Field}
 The daily values of
 the components
of the interplanetary magnetic field $B$ and 
$B_x$,  $B_y$ 
in 2003 with one day step are plotted in  Fig. 5.
The auto-correlation analysis (and FFT as well while not shown in this paper)  reveals a  
27-day periodicity in the $B_x$ and $B_y$ components,
shown in the bottom plot of Fig. 6, 
a 29-day periodicity in the $B_z$
component, and a 26-day periodicity in the variability of the
total value of solar wind magnetic field $B$. 

 These periodicities  can be considered as
 an evidence that  the $B_x$ and $B_y$ components have an origin 
 in the solar middle latitudinal zones rotating relatively fast.
 27 days correspond  to the sidereal period of the solar magnetic
 field rotation  on the  latitudes  of about $\pm 40$ degrees,
 while the sidereal period of $SMF$ rotation reaches 29 days
 on the heliographic latitudes  of about $\pm 60$ degrees
 as it can be seen in  Fig. 12 in  Gavryuseva  (2006).
 Perhaps  $B_z$ component is mainly sensitive to the fast wind flow
 from the polar coronal holes.

 A clear anti-correlation between the  $B_x$ and  $B_y$ components
 of the $IMF$ is illustrated in Fig. 5 for the daily data taken in 2003.
 On the upper plot the $B_x$ and  $B_y$ components are shown by
 continuous and dotted lines respectively.
  The correlation between them with zero shift is equal to -0.85,
  and the period of cross-correlation is $27.00\pm 0.38$ days
  as  seen in the bottom plot in Fig. 6.
  This anti-correlation is due to
  the orientation of the coordinate system of the interplanetary magnetic field
  where the $X$ is directed Sun-ward.
\section{Short Term Variability of Interplanetary parameters}
   The  solar magnetic field,  in addition
   to the $4$-zonal structure with the 22-year periodicity,
   presents a clear topology
   consisting of the $SMF$ Waves Running  through Latitudes  ($WRL$-structure)
   with a quasi 2-year periodicity  shown in the bottom plot
   of Fig. 1 and discussed in details by
   Gavryuseva,  (2005, 2006, 2006 a,b,c).
   This behaviour should be compared with
   the variability of the  $IMF$, 
   solar wind and geomagnetic data on a short term scale.

  We first applied the filter of long-term periodicities
  to the $SW$ sets of OMNI data,
  then the shortest variabilities,
  for example those caused by the orbital rotation, were filtered by
  1-year smoothing. Such smoothed residuals of the OMNI data
  are called Filtered OMNI  or Filtered Solar Wind (FSW) data
  and include the residuals of the interplanetary magnetic field,
  solar wind and geomagnetic parameters.

   Short-term changes of the $FSW$ are  plotted
   in Fig. 7 for the following parameters:
   $B$, $B_x$, $B_y$, $B_z$;
   $T_p$, $V_p$, $E$,  $N_{\alpha}/N_{p}$;
   $P$, $N_p$, $N_{\beta}$, $M_a$;
   geomagnetic indices $AE$, $K_p$, $DST$ and Sunspot number $SSN$
   (from the top to the bottom rows).
   A periodicity of 2--3-year 
   is clearly seen in the temporal behaviour of these quantities.

  The $SW$  and geomagnetic variability is  discussed  by different authors
  (see for example,  Kane,  (2005a, 2005b);
    Rivin, (1989) and references there). Fraser-Smith (1973) investigated the
  27-day variation of geomagnetic activity. Clua de Gonzalez et al. (1993) found  that the 6-month periodicity has a multiple origin, 
and 4-year in the monthly Ap power spectrum is associated to the double peak 
structure observed in the geomagnetic activity.
It was noted by  Kane, (2005b) that solar indices have a
  quasi-biennial oscillation in form of
  double peaks separated by 2-3 years   during sunspot cycle.

  Rivin  (1989)  found that the amplitude of the biennial variation
  of the geomagnetic field is modulated by a 10-year period.
  This well agrees with the presence of the $WRL$-topology
  in the photospheric magnetic field, whose strength varies 
  with the 2-year periodicity and  interferes with the
  4-zonal  structure that changes with the 22-year period.
  The detailed discussion of the periodicities of the $IMF$, 
  solar wind and geomagnetic activity is not a subject of this paper,
  and therefore we do not present here the table of the $SW$ periodicities.
  We give only an illustration of a short term variability
  of the OMNI data  in Fig. 8.  
   The correlation coefficients  were
   calculated for a 4-year long subset of the $FSW$ data running
   through a $FSW$ residual calculated as a difference
   of an experimental data set and 4-year running means
   of the following parameters:
   $B$, $B_x$, $B_y$, $B_z$;
   $T_p$, $V_p$, $E$,  $N_{\alpha}/N_{p}$;
   $P$, $N_p$, $N_{\beta}$, $M_a$;
   geomagnetic indices $AE$, $K_p$, $DST$ and  $SSN$
   (from the top to the bottom rows in Fig. 8).

 The   $B_x$,  $B_y$, $B$ values have
 a 378-day periodicity.
 This period is not equal to one year (365.25 days)
 (related to the orbital rotation of the Earth)
 due to the rotation of the Sun.
 The intensity of the interplanetary magnetic field $B$ exhibits
 a half a year and  a 1.5-year periodicities.
 $B_z$ component has a 1.35-year periodicity.

   A quasi biennial periodicity is seen
   in  almost all the OMNI data sets.
   The plots of Fig. 8
   confirm the presence of periodicities of about 2-3 years
   in the $IMF$, 
   solar wind and geomagnetic  parameters
   in agreement  with the results
   of other  studies  Rivin, (1989).
 \section{ Relationship Between Photospheric
  and Interplanetary Magnetic Fields in 2003}
 First we concentrate on the relationship between
 the  photospheric magnetic field and interplanetary  magnetic field.
 This helps to establish the channel of the Sun-Earth connection.

 Daily data for 1996, 2000 and 2003 have been used to
 examine the delay between the phenomena on the Sun and 
 on the Earth orbit. Here we present the result for the data sets taken in 2003.

 The   correlations  between the $SMF$ and $IMF$ in 2003
 provide the possibility to find their relationships during
 a period of high solar activity, after the polarity inversion in 2001
 in the sub-polar regions
 and to determine the optimal delay between the $SMF$ and $IMF$.

  In Fig. 9  from the top to the bottom 
  the correlations
  between the solar photospheric field 
  measured in 2003 and  the $IMF$ field intensity $B$, 
  the $IMF$ components
  $B_x$, $B_y$ and $B_z$
  as functions of time lag in days
  and of heliographic latitude are presented.

  Yellow and red (blue and green) colors indicate
  positive (negative) correlation coefficient values.
  The contours correspond to zero level and to the levels
  of $\pm 0.5$ of the maximum value of the correlation coefficient.
 
  The solar rotational periodicity is visible on all the
  plots of Fig. 9 in agreement with the
  solar origin of the interplanetary magnetic field.
  These plots show that the interplanetary field is
  sensitive to the magnetic field of the Sun
  with an optimum delay of about 4 days.
  This delay is used to study the the correlation
  between the sets of the Bartels means of the $SMF$ and OMNI data.
 \section{ Photospheric and Interplanetary Magnetic Fields
           through Cycles of Solar Activity}
  In order to analyse the  long term relationship between the photospheric
  and the interplanetary field
  over the last three cycles of the sunspot activity,
  the correlations between 2-year long
  sub-sets of the $SMF$ and  $IMF$ data running in time with 
  1 Carrington Rotation (CR, 1 CR = 27.2753 days) step  have been calculated.
  In Fig. 10  the correlation coefficients between
  the $SMF$,  intensity $B$  and components $B_x$, $B_y$ and $B_z$ 
  of the $IMF$  are plotted  from  the top to the  bottom.
 Fig. 10 gives a very interesting panorama
 of the relationships between  the photospheric
 magnetic field and the $IMF$ over the last three cycles of activity.
 Firstly, the correlation between the $SMF$ and  $B_x$ component,
 and between the $SMF$ and $B_y$ component is of opposite sign
 on a long-time scale.
 This fact could be used to verify and to confirm conclusions
 related to one of them.

Secondly, as immediately follows from the top and middle plots,
a strong correlation (anti-correlation)  does exist 
between the photospheric  $SMF$ 
on the latitudes above the 50 degrees
and $B_y$ ($B_x$) component  during the minima of solar activity
which took place around 1976-1977, 1985-1986 and 1995-1996.
The correlation reaches the level of 0.92.
Contribution of the sub-polar regions to the solar wind near the ecliptic
was significant when the high latitude field was very strong during the minima of activity.
In 1996   the  correlation (anti-correlation) was particularly strong 
between  the photospheric field  over all the latitudes
and $B_x$ and $B_y$ components.
In this period the solar magnetic field in both hemispheres
was rising up.
Magnetic field was positive almost everywhere from 50 degrees   
in the southern hemisphere to the pole of the northern hemisphere.

We could conclude that during periods of quiet Sun the generation of
the interplanetary field takes place mainly in the sub-polar regions
while the middle and low latitudinal zones could also contribute 
synchronously. These results agree with conclusions of other studies
  (see, for example   Gonzalez et al.,  (1994, 1999);
   Wang et al.,  (2000);  and references therein).

 The $B_z$ component correlates to the pre-equatorial $SMF$ 
 in the northern hemisphere and middle latitude zone
 in the southern hemisphere during maximum of cycle No 21.
 It is opposite to its behaviour in cycle No 22 and similar in
 cycle No 23, but, additionally, at the maximum of the last cycle
 $B_z$ correlates to the sub-polar $SMF$ 
 in both the northern and  southern hemispheres.

 Generally speaking, the interplanetary magnetic field was well synchronized with the
 photospheric field variability in the broad region of latitudes in 
 cycle No 23;  this means that the low-latitude solar wind originated from
isolated low-latitude and mid-latitude coronal holes could contribute
significantly during the cycle.
This finding contradicts the paradigm of a
solar wind source consisting of two polar outflows only
flanking a planar current sheet (Pizzo, 1982), but is in
agreement with the results of  Wang et al. (2000) demonstrating
that around the solar minimum the sources of the Sun's open magnetic field lines
whose extension constitutes the $IMF$ are the big polar coronal holes,
while at solar maximum these are the small low-latitude coronal holes.

\section{Relationships between the SMF and OMNI /
         Data on Long- and Short-Term Scales}
 The study of the general relationships between the Sun, solar wind
 and geomagnetic perturbations can be performed by comparison
 of the variability of the solar magnetic field at different latitudes with
 the temporal behaviour of the OMNI data
 over several activity cycles.
 Correlation coefficients $K_{cor}$  between
 full data sets from 1976 to 2004 with one Bartels rotation resolution of
 the photospheric field from one side and of
 the solar wind and geomagnetic parameters ($SW$-sets)
 on the other side have been calculated 
 with the shift up to 22 years.
 
 Fig. 11 shows these correlation coefficients as functions of time-shift
 (in years) and latitude for  $K_{cor}$ between
 the yearly means of the $SMF$ and of 
   the interplanetary field intensity $B$ and
   $B_x$, $B_y$, $B_z$ components  (upper plot),
   and then down  to the bottom for  $K_{cor}$ between
   the yearly means of the $SMF$ and
   $T_p$, $V_p$, $E$,  $N_{\alpha}/N_p$ ratio;
   $P$, $N_p$, $N_{\beta}$, $M_a$;
   $AE$, $K_p$, $DST$   and  $SSN$.
   The data used take into account the sign of the parameters.
   The $K_{cor}$ presented in Fig. 11 was smoothed
    by a 1-year running window.
   The "zero" shift corresponds to the $SW$ optimum   delay of 4 days
   discussed in section 3.

   The latitudinal dependence
   of these correlations is a consequence of the 4-zonal $SMF$ topology.
   The correlation between the $SMF$ and  $B_x$, $B_y$ and $B_z$ components
   has a 22-year periodical behaviour due to 
   the 22-year variability of the solar magnetic field.
  The correlation coefficients $K_{cor}(SMF, B_y)$  between
  the $SMF$ and $B_y$ is opposite to the
  correlation coefficients $K_{cor}(SMF, B_x)$ and
  $K_{cor}(SMF, B_z)$  between the $SMF$ and
  $B_x$ and $B_z$ components.
  The relationship between the photospheric field and $E$ is
  similar to the one between the $SMF$ and $B_y$,
  while the $K_{cor}(SMF, T_p)$ and $K_{cor}(SMF, V_p)$ are similar to
  the $K_{cor}(SMF, B_x)$, and opposite to the  $K_{cor}(SMF, B_y)$,
  while $K_p$ has the same relationship with the $SMF$ as
  $B_x$ component of the $IMF$ and opposite to the $K_{cor}(SMF, DST)$.

   The correlation between the $SMF$ and the intensity $B$
   shows a quasi 30-year periodicity.
   The relationship between the photospheric field and
   $P$, $N_p$, $N_{\beta}$,  $M_a$ is similar, while
   $K_{cor}(SMF,  N_{\alpha}/N_{p})$ has an opposite sign.
   The correlation coefficient
   $K_{cor}(SMF,  AE)$ is similar to the $K_{cor}(SMF, B)$.

   Figure 11 illustrates the periodic character
   of the relationships between the $SMF$ and the $SW$ data.
   As it was deduced the $SW$ data can be divided into two groups
   having similar dependence on the time-shift with the photospheric field.

   For search on the $SMF-SW$ dependence it is important to
   analyse  the correlation between them with the "zero" shift
   which corresponds to the optimum delay.
   Figure 12 shows the coefficients of correlation
   between the 1-year means of the full data sets of the  photospheric field
   at different latitudes and the interplanetary magnetic field,
   solar wind and geomagnetic parameters.
   The latitudinal dependences of the  $K_{cor}(SMF, SW)$
   permit to select two groups of the $SW$ parameters which have
   similar sensitivity to the $SMF$ at different helio latitudes.
   The first group is composed of
   $B$, $P$, $N_p$, $N_{\beta}$, $M_a$,
   $-N_{\alpha}/N_{p}$, $AE$ and -$SSN$.
   The second one includes
   $B_x$, $-B_y$, $B_z$, $T_p$, $V_p$,
   $-E$, $K_p$ and $-DST$.
   Figure 12  confirms the presence of two groups as deduced from Fig. 11.

   The positive value of   $K_{cor}$ is an indicator of 
   a   possible  connection between the $SMF$ and a $SW$, 
   but it is not enough to
   conclude that the $SMF$ and a $SW$ are physically dependent.

   Since biennial variability is  present  in
   the  solar magnetic field 
   it is necessary to verify the   relationship
   between the variability of the  $SMF$, $IMF$, 
   solar wind and geomagnetic data on a short-term scale
   of about 2 years.
   This is also useful for the study of a $SMF-SW$  physical dependence.
   Such an analysis of the relationship of the  short-term variability
   of the solar  magnetic field and $FSW$  data 
   has been performed to complete this study.
   The letter $F$ is added before the abbreviation of the title
   of a parameter to indicate that the residuals were calculated to
   filter the long-term variabilities of the corresponding parameter.
   The latitudinal dependences of the  $K_{cor}(FMF, FSW)$
   were found  for the correlation between the residuals of
   the photospheric magnetic field  (FMF) with the residuals
   of the  different  $SW$ data ($FSW$)
   calculated as the difference between yearly and 4-year means
   of the $SMF$ and $SW$ data.
   
   Coefficients of correlation  with the optimum delay
   between the short-term variabilities
   of the photospheric field at different latitudes and
   the interplanetary magnetic field,
   solar wind and geomagnetic parameters 
   are plotted in  Fig. 13.
   The same two groups could be selected in the OMNI data
   from the point of view of their relation with the photospheric field.
    These relationships are better illustrated in Figs. 14 and 15
    where the  latitudinal dependences of the  $K_{cor}(SMF, SW)$
    for long-term variability (plots on the left) and of 
    the $K_{cor}(FMF, FSW)$
    for short-term variability (plots on the right) are shown for 
    the following $SW$ data:\\
    $B$, $P$,  $N_p$, $N_{\beta}$, $M_a$,
    $AE$, $SSN$ (Fig. 14);\\
    $B_x$, $B_y$, $B_z$ components,
    $V_p$, $T_p$, $E$,  $N_{\alpha}/N_{p}$,
    $K_p$, $DST$ (Fig. 15).

   In this way it is established that
   there are two groups of the  interplanetary data
   led by  $B$ and  $B_x$ (or -$B_y$, $B_z$)
   which respond to the
   variability of the photospheric magnetic field
   in a similar way
   (from the point of view of the periodic character and
   of the latitudinal dependence of the $K_{cor}(SMF, SW)$ or
   the $K_{cor}(FMF, FSW)$
   and they have a similar response in the
   geomagnetic perturbations of $AE$, $K_p$ and $-DST$ indices).
    
   \section{Some summary remarks}
  Southward-directed interplanetary magnetic field is considered
  a primary cause of geomagnetic perturbations
  (Durney, 1961;  Gonzales et al., 1994, 1999).
  As  a consequence the orientation of the interplanetary magnetic field
  (Axford and McKenzie, 1997;  Low, 1996;
    Parker, 1997;  Smith, 1997)
  plays an important role.

  The solar activity phenomena depend on the sunspot cycle, which can be
  characterized by the variability of the $SMF$ intensity
  in time and along the latitudes.  The topology of the solar
  magnetic field influences the geomagnetic perturbations
  through the intensity and orientation of the interplanetary magnetic
  field and/or through other parameters of the solar wind.
  In this approach we could understand the presence of two groups of
  the OMNI data similarly sensitive to the basic topology of the magnetic
  field of the Sun (from the point of view of the dependence on latitude 
  and phase-shift of the correlation of the coefficients
  with the mean latitudinal magnetic field).

  The formal and complete study of the problem of solar-terrestrial relations 
has been performed and the connections between the processes on the way
 from the Sun to the Earth have been revealed. A useful information was 
deduced from the temporal behaviour and dependence of the correlation 
of the photospheric magnetic field and different parameters of interplanetary
 space and geomagnetosphere.

It was revealed directly from the experimental data that there are 
two groups of $SW$ parameter which respond in a similar way to
the behaviour of solar characteristics.
  We found that the photospheric field
  influences  the magnitude of the interplanetary field
  and, in the same way, the proton density, flow pressure,
  Alfven Mach number and plasma $\beta$ respond to the $SMF$.
  Moreover the $AE$-index behaves in a similar way as the above mentioned
  solar wind parameters.

  On the contrary, regarding  the planetary geomagnetic activity index $K_p$
  we can deduce that solar activity events  (CME, magnetic field intensity, 
sunspots, etc.)
  through perturbations of
  the $B_z$ component ($B_x$, $B_y$ components) of the $IMF$, 
  the   proton temperature $T_p$, plasma  speed $V_p$,
  $N_{\alpha}/N_{p}$ ratio influence the $K_p$ index.
  The variations of the $-B_z$ ($B_y$) component
  produce the perturbations
  of the $DST$ index, and they are of opposite sign of  the $K_p$
  and $B_x$ time   dependence.

   It was also revealed  from the experimental data that the solar magnetic fields and 
solar activity  processes  originated bellow $\pm 55$ degrees  propagate up  to the 
Earth orbit and produce the perturbations of the magnetosphere
   (Gavryuseva, 2006 c,f; 2008b,  Gavryuseva \& Godoli, 2006).

 These results are useful for understanding
 the origin of solar wind and geomagnetic perturbations
 and for long-term predictions.

 \section*{Acknowledgments}
      I thank the WSO and OMNI teams for making available
      data of measurements of the solar magnetic field,
      solar wind and geomagnetic quantities.
    I am grateful to Prof. G. Godoli  for his stimulating interest in these results
    and Profs. B. Draine, L. Paterno and E. Tikhomolov  for  help in polishing this paper 
    and useful advises.

\newpage
\clearpage
 \begin{figure}
  \centerline{
  \includegraphics[angle=90, width=39pc]
    {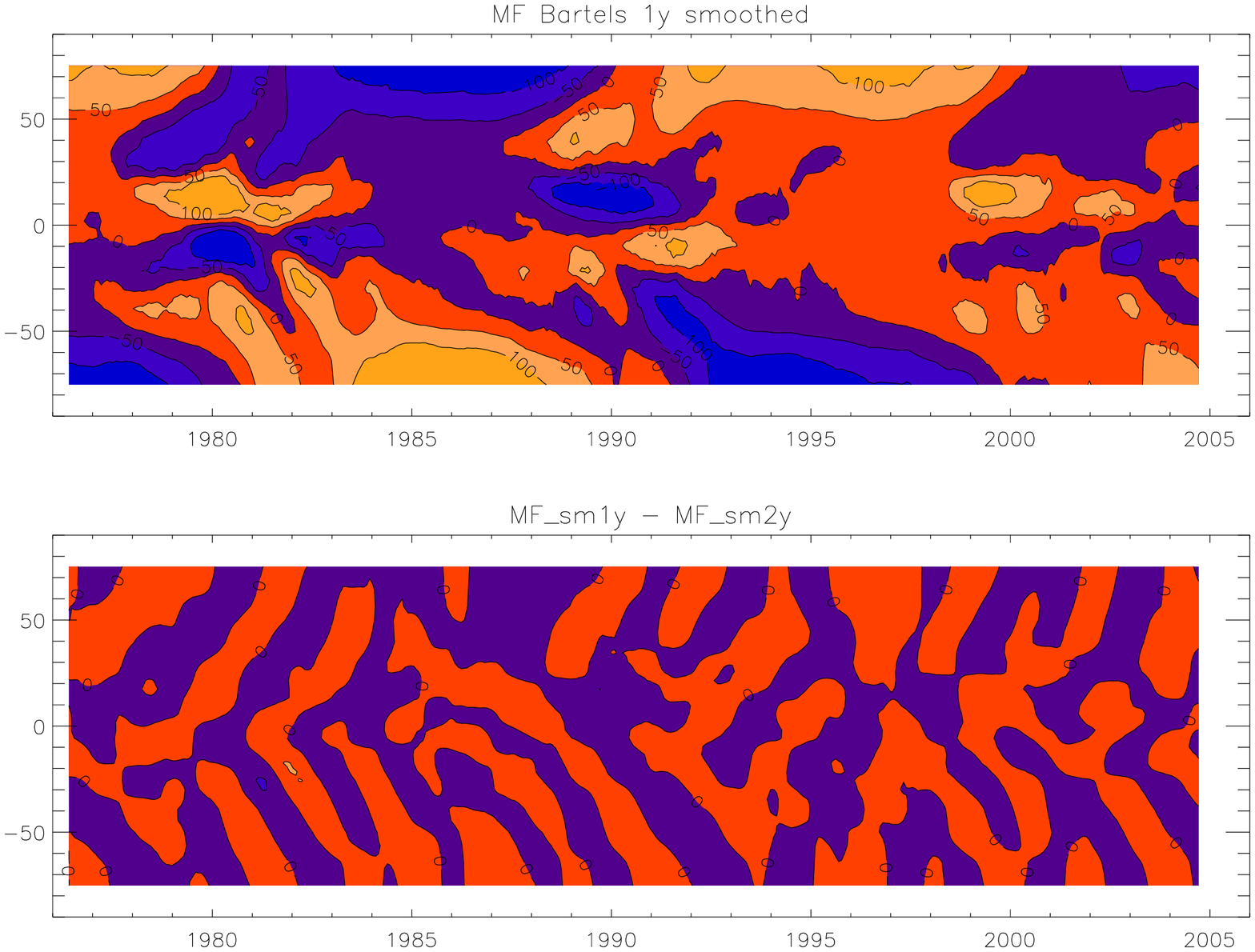}
    }
  \caption{
  The  1-year mean photospheric field (upper plot)
  and its short-term variable part (bottom plot)
  as a function of time and  latitudes. Red (dark blue)  and orange (light blue) colors 
indicate positive (negative) intensity values. 
The contours correspond to the 0, +/-50, +/-100 micro Tesla.
       }
  \end{figure}
\clearpage
 \begin{figure}
  \centerline{
  \includegraphics[angle=90, width=39pc]
    {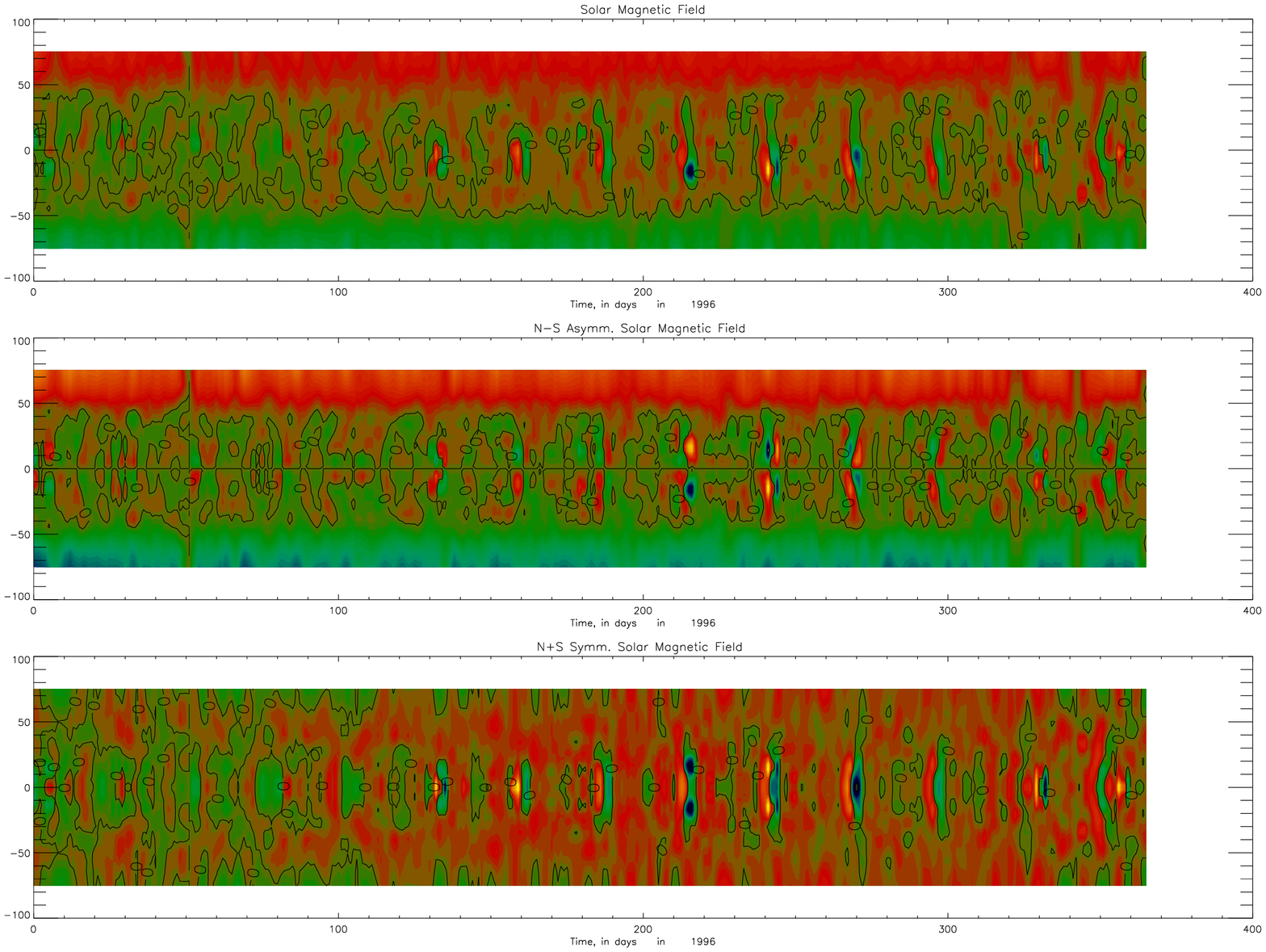}
    }
  \caption
 { Distribution in latitude and in time with 1 day
  step of the original WSO data (upper plot),
  the North-South antisymmetric part of the WSO  data (middle plot)
  and the North-South symmetric part of the WSO data (bottom plot)
  of the solar magnetic field   in 1996.
  Red (green) colors indicate
  positive (negative) correlation coefficient values.
  }
  \end{figure}
 \begin{figure}
  \centerline{
  \includegraphics[angle=90, width=39pc]
    {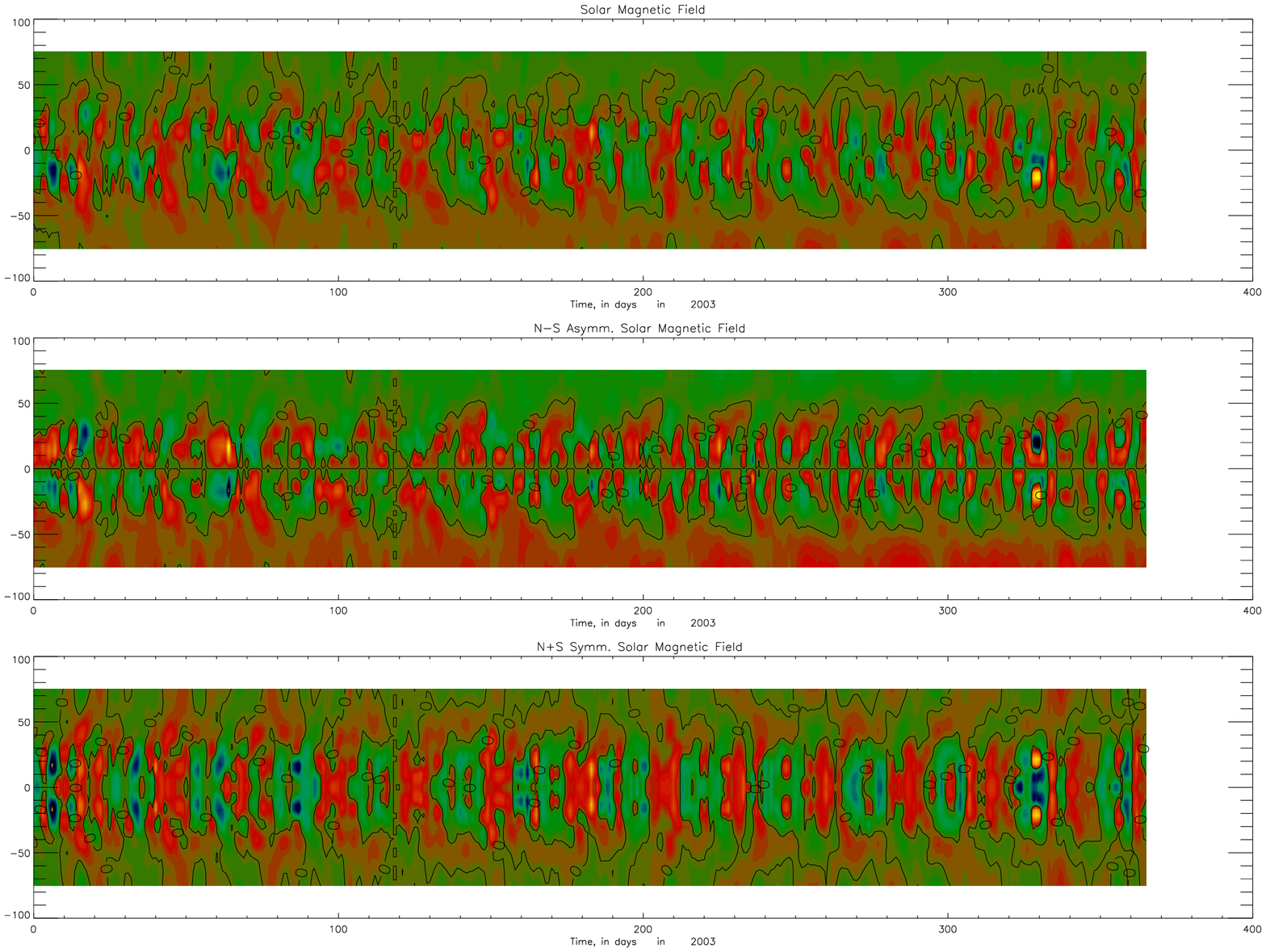}
 } 
 \caption
 {  Distribution in latitude and in time with 1 day
  step of the original WSO data (upper plot),
  the North-South antisymmetric part of WSO  data (middle plot)
  and the North-South symmetric part of WSO data (bottom plot)
  of the solar magnetic field   in 2003.
  Red (green) colors indicate
  positive (negative) correlation coefficient values.
  }
  \end{figure}
\clearpage
 \begin{figure}
  \centerline{
  \includegraphics[angle=90, width=39pc]
  {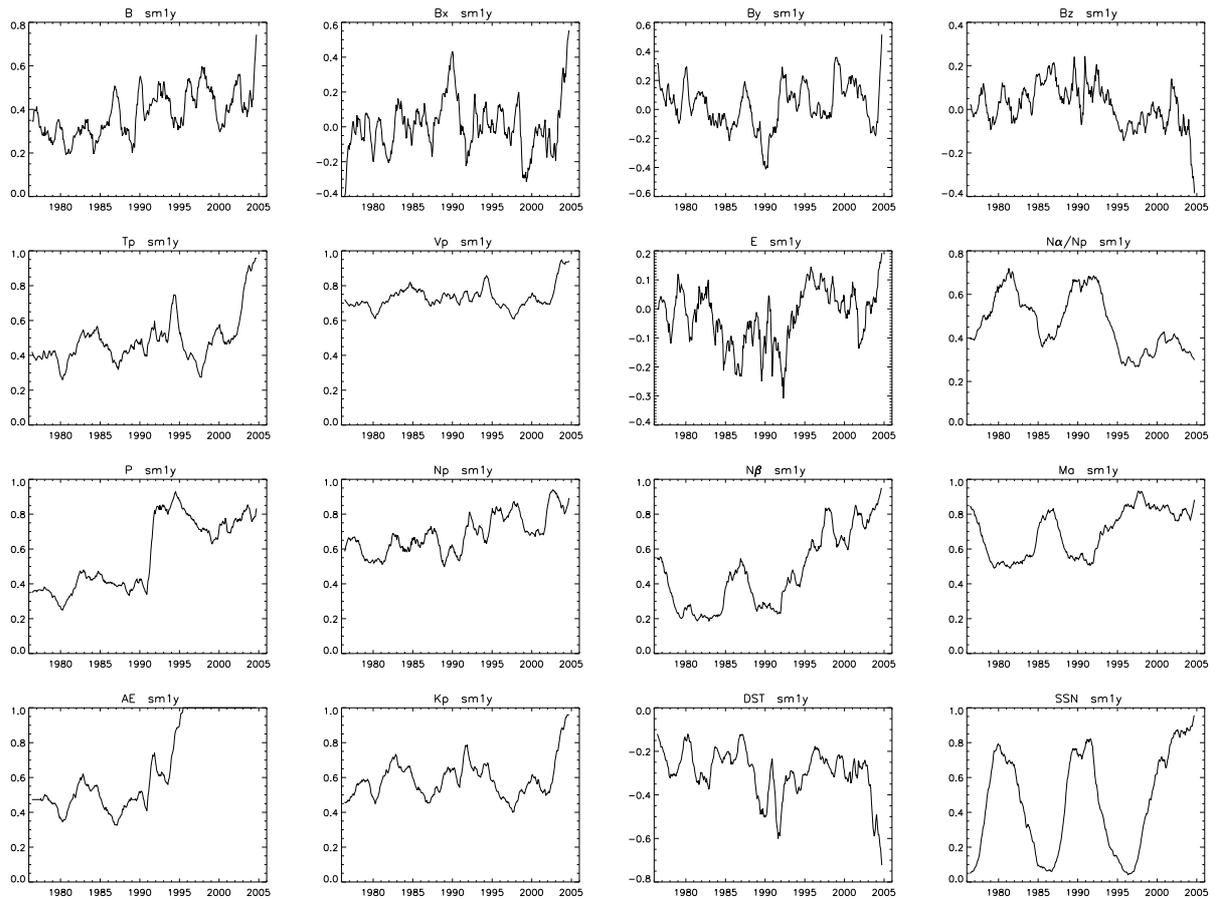}
  }
  \caption
 {Total intensity and components $B_x$,  $B_y$, $B_z$
 of the interplanetary magnetic field, solar wind
 and geomagnetic parameters from the OMNI data bank
 with 1 Bartels rotation resolution. These data are normalized to the maximal value 
and then smoothed by 13 Bartels rotations (about 1 year).
 }
  \end{figure}
\clearpage
 \begin{figure}
  \centerline{
  \includegraphics[angle=90, width=39pc]
    {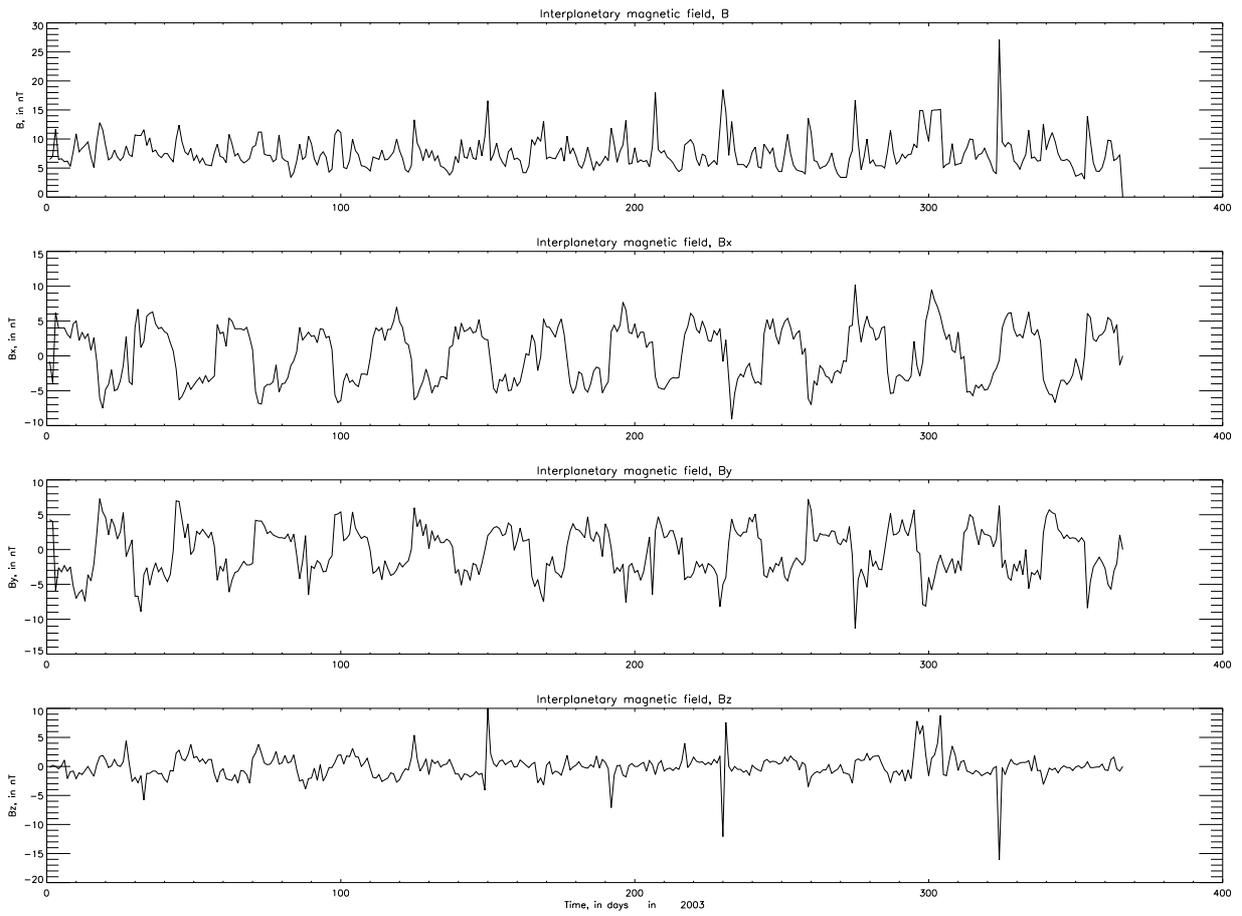}
  }
  \caption
 {Total intensity B and components $B_x$,  $B_y$, $B_z$
 of the interplanetary magnetic field in nT
 in 2003 with 1 day step.
 }
  \end{figure}
\clearpage
 \begin{figure}
  \centerline{
  \includegraphics[angle=90, width=39pc]
      {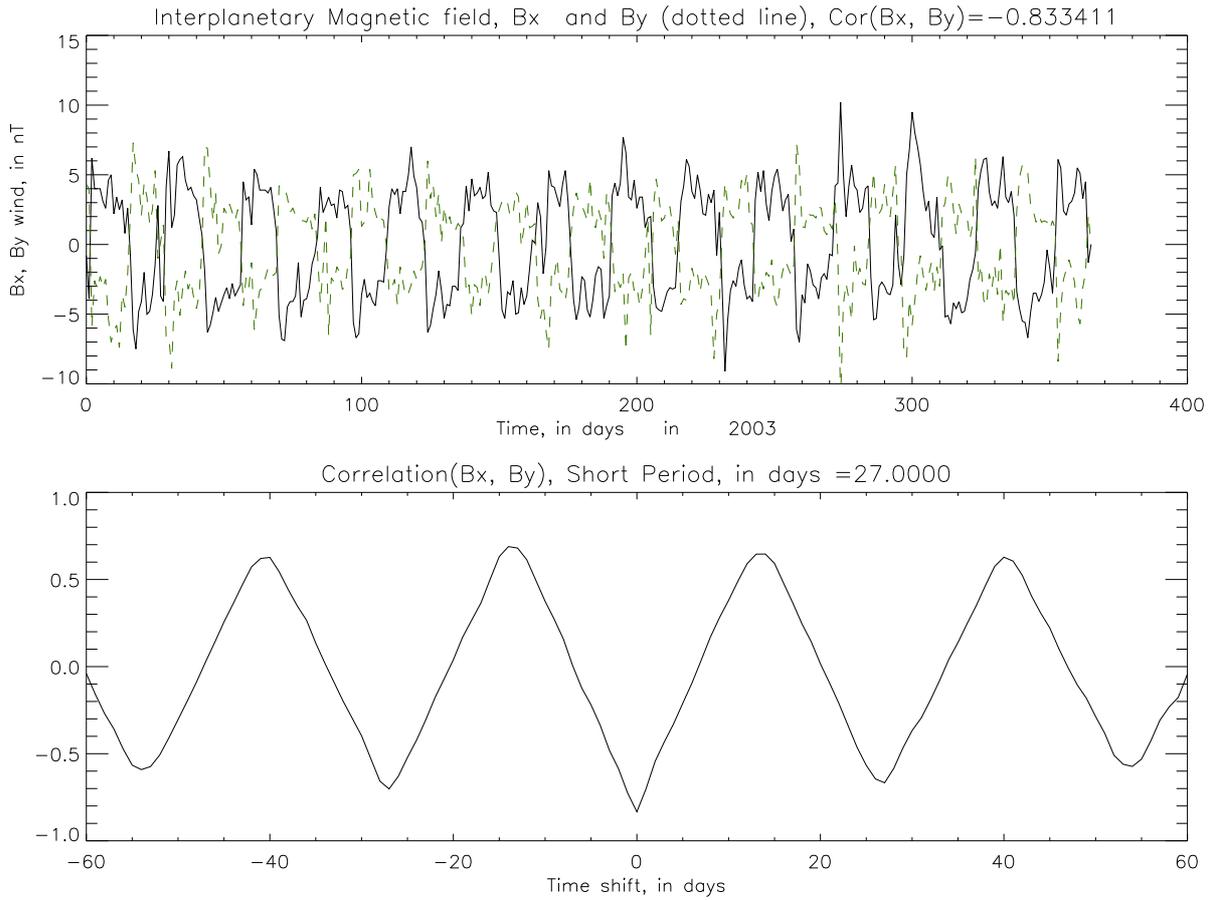}
  }
  \caption{
 On the upper plot the $B_x$ and  $B_y$ components of the 
 solar wind magnetic field in 2003
 are shown by continuous and dotted lines.
 The coefficient of the cross-correlation between them as a function of the shift in days is presented ($X$ axis)
on the bottom plot.
  }
  \end{figure}
\clearpage
 \begin{figure}
  \centerline{
  \includegraphics[angle=90, width=39pc]
    {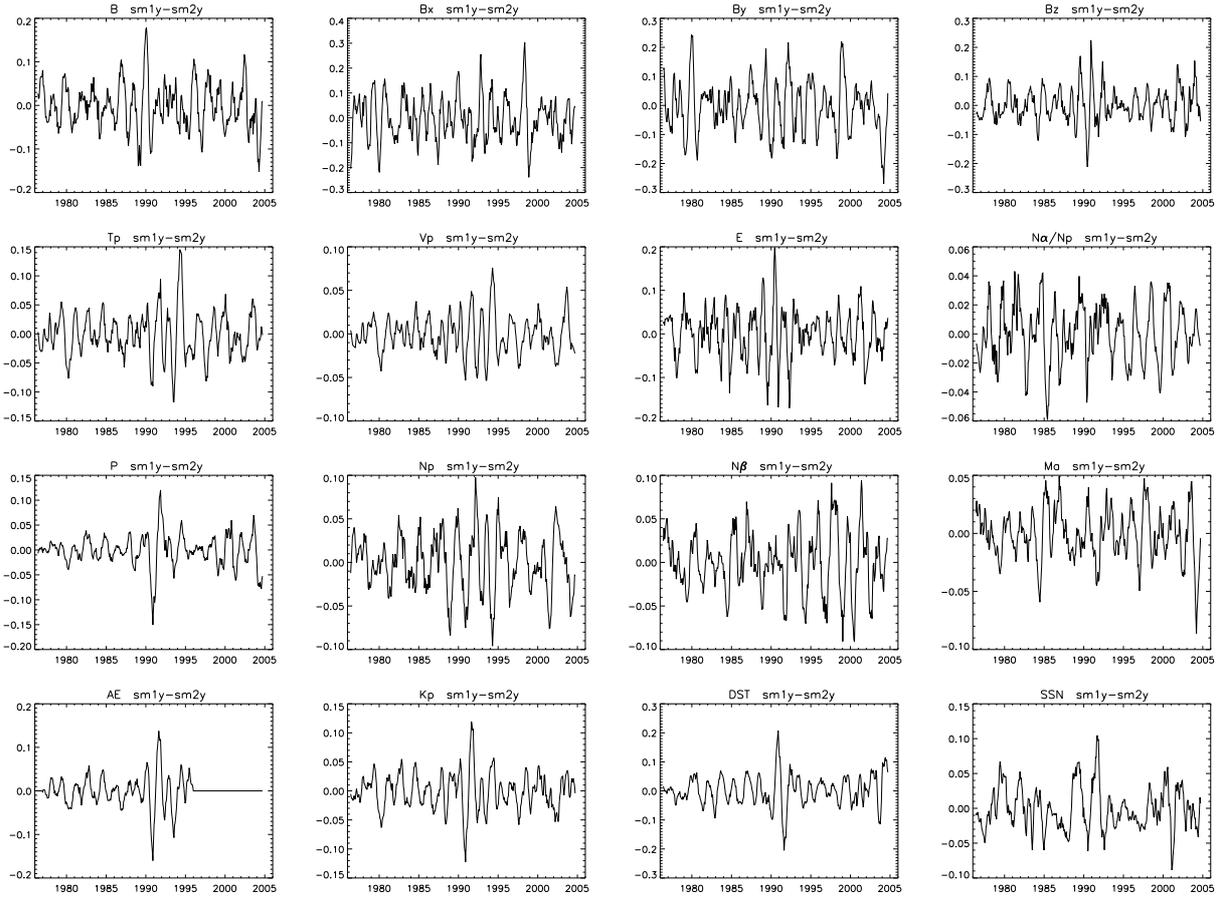}
    }
  \caption{
The residuals of the  interplanetary magnetic field, solar wind and geomagnetic parameters 
through solar cycles (the filters of periods longer than 4 years and shorter than 1 year 
have been applied). Short term variability of the interplanetary magnetic field, 
solar wind and geomagnetic parameters through solar cycles.
}
  \end{figure}
\clearpage
 \begin{figure}
  \centerline{
  \includegraphics[angle=90, width=39pc]
    {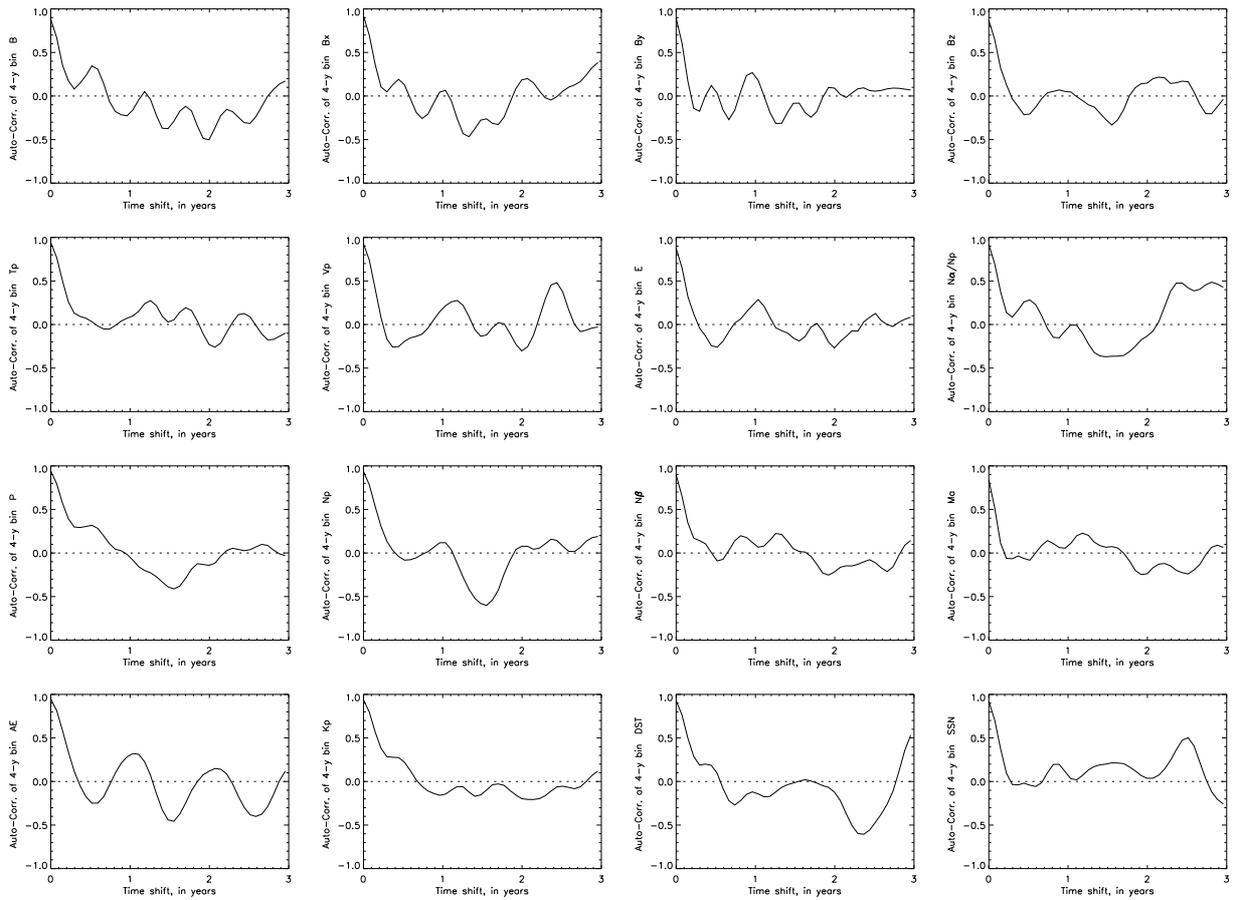}
    }
  \caption{
  Auto-correlation coefficients demonstrate short term periodicities of about 1- and 2-year long for
  the interplanetary magnetic field,
  solar wind and geomagnetic parameters.
 }
  \end{figure}
\clearpage
  \begin{figure}
  \centerline{
  \includegraphics[angle=90, width=39pc]
    {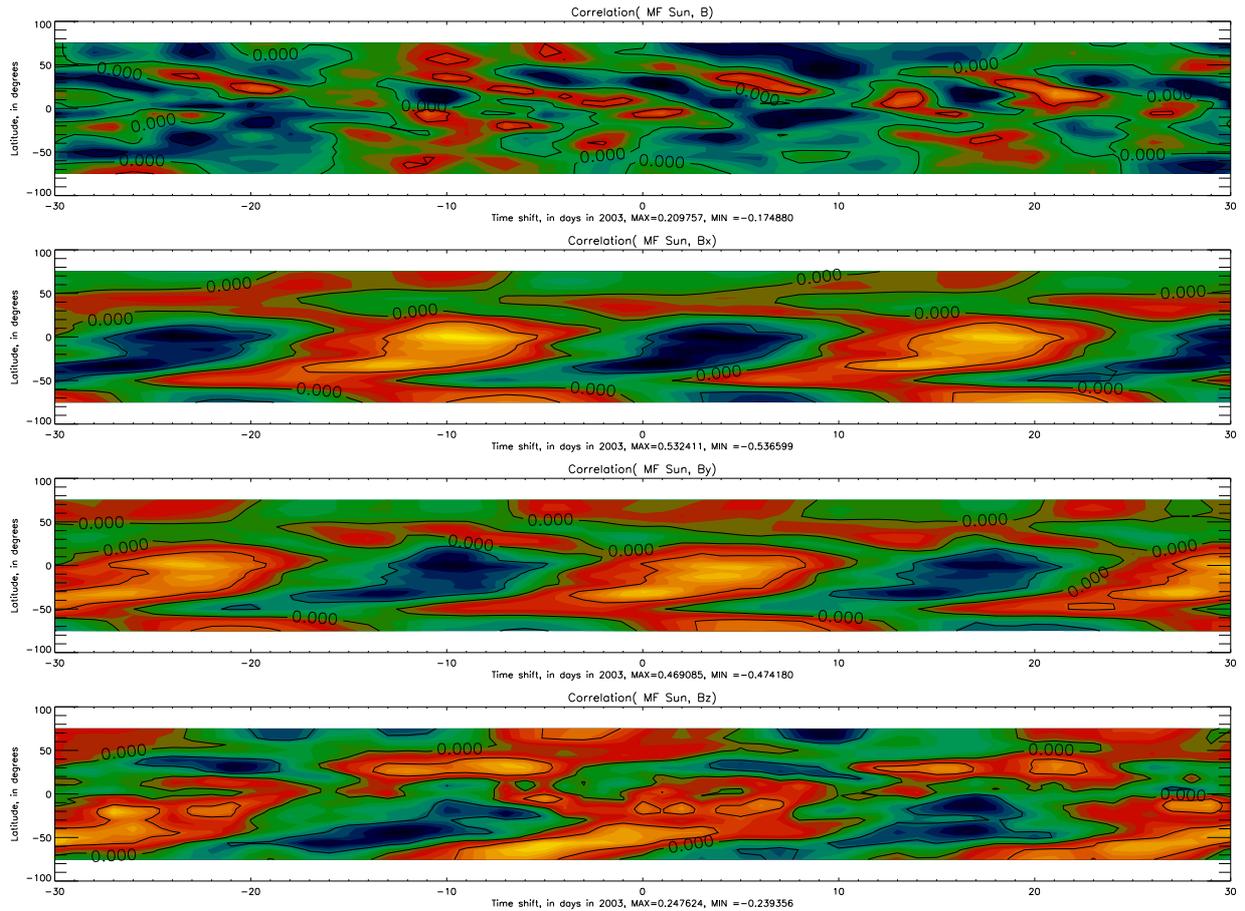}
  }
  \caption{
  Correlation as a function of time lag in days between the
  original photospheric field  and
  (from the top to the bottom)
  the total intensity $B$,
  the  components  $B_x$, $B_y$  and $B_z$  of the
  interplanetary magnetic field.
  Yellow and red (blue and green) colors indicate
  positive (negative) correlation coefficients of high and 
  lower levels of positive (negative) values.
 }
  \end{figure}
\clearpage
 \begin{figure}
  \centerline{
  \includegraphics[angle=90, width=39pc]
    {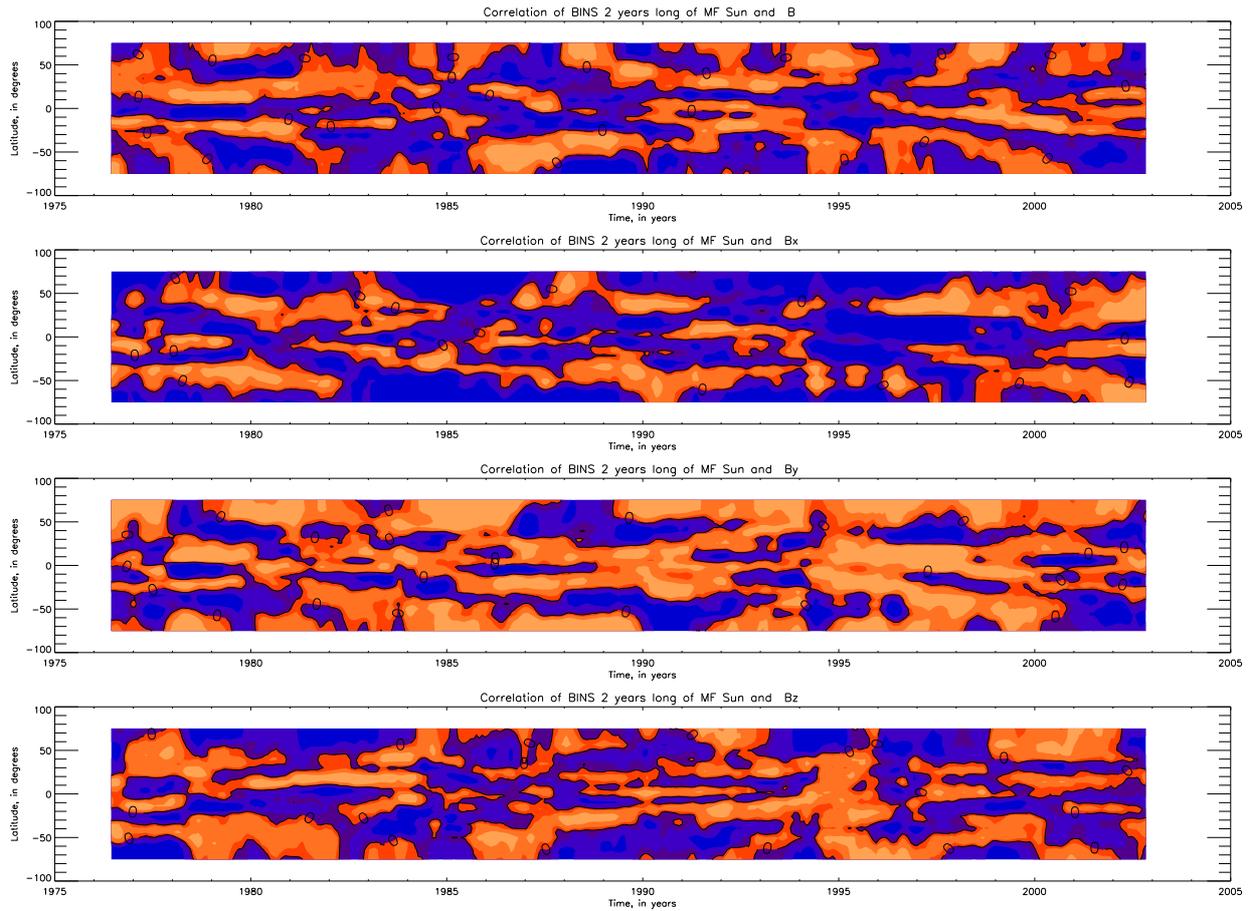}
  }
  \caption{
   Correlation between the 2-year long sub-sets of the photospheric field
   at different latitudes ($Y$ axis)
   with the interplanetary magnetic field intensity  $B$  and the 
   component  $B_x$ (upper plot),
   and the components  $B_x$, $B_y$ (middle plots) and $B_z$ (bottom plot)
   as a function of time and latitude.
   Yellow and red (blue and green) colors indicate
   positive (negative) correlation coefficient values.
 }
  \end{figure}
\clearpage
  \begin{figure}
  \centerline{
  \includegraphics[angle=90, width=39pc]
    {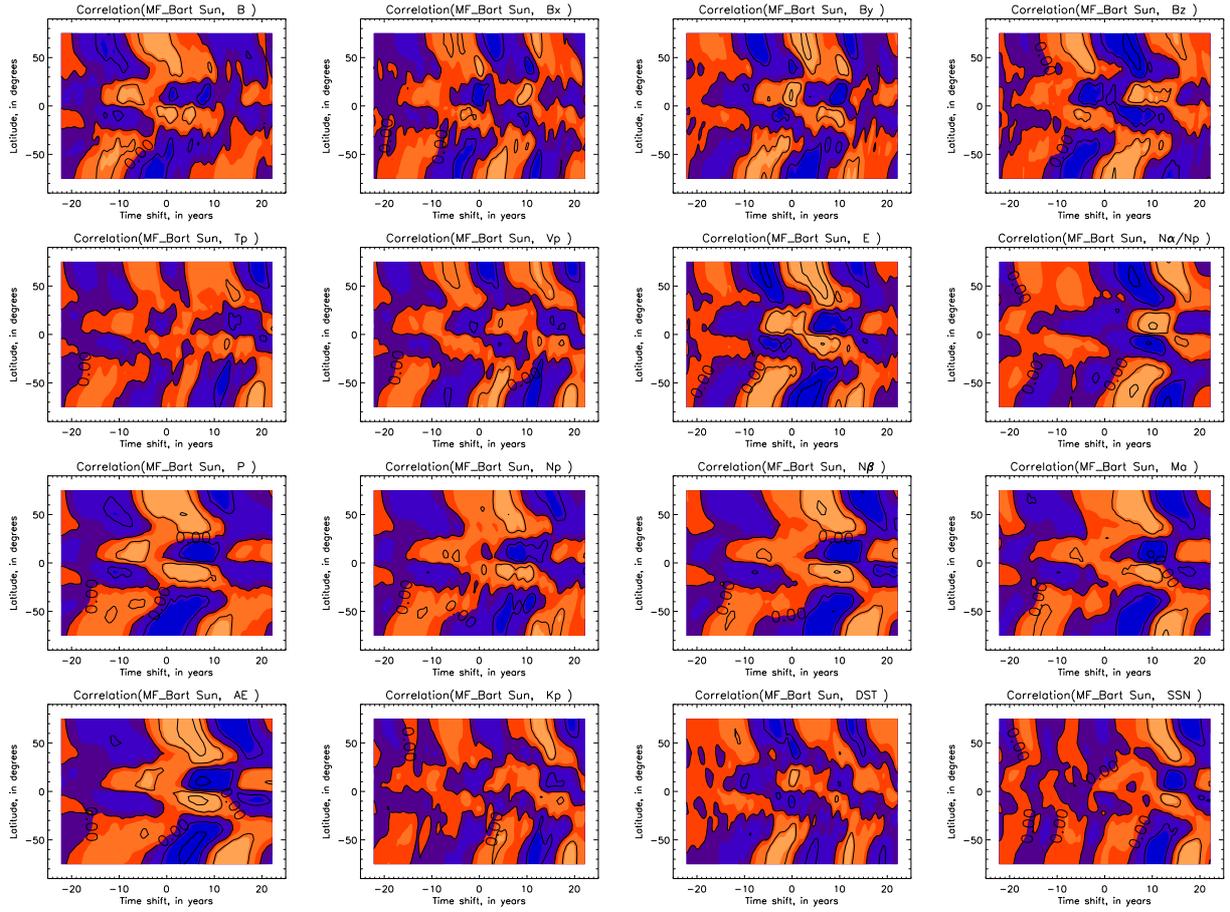}
  }
  \caption{
  Correlation coefficients  of the 1-year mean 
 photospheric field (taking into
  account its polarity) at different latitudes ($Y$ axis)
  with solar wind parameters
  and geomagnetic perturbations
  as a function of delay in years ($X$ axis) and  of latitude.
  Orange and red (blue) colors indicate
  positive (negative) correlation coefficient values.
 }
  \end{figure}
\clearpage
  \begin{figure}
  \centerline{
  \includegraphics[angle=90, width=39pc]
    {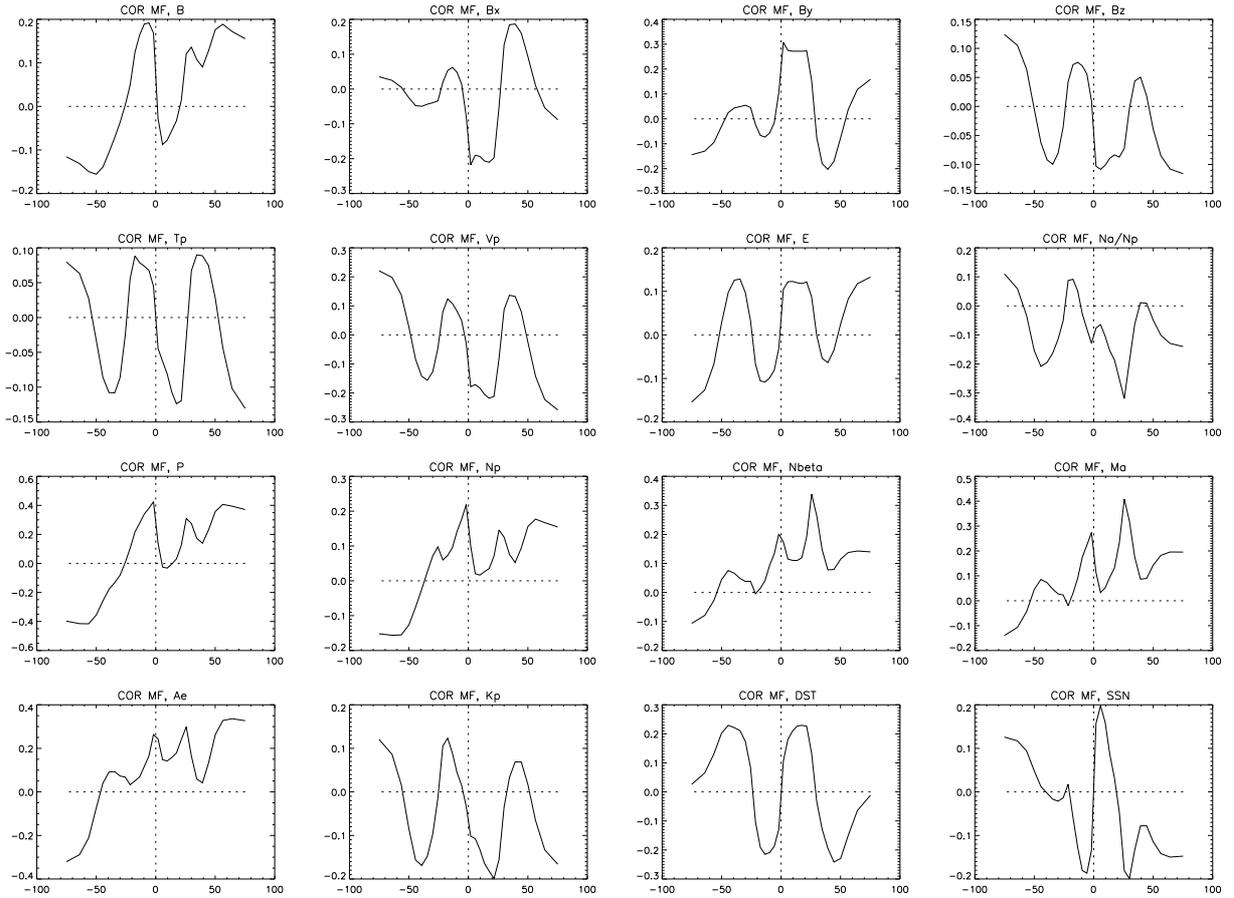}
 }
 \caption{
  Coefficients of correlation between  1-year mean photospheric field
  at different latitudes and
  the interplanetary magnetic field,
  solar wind and geomagnetic parameters  with the fixed delay of 4 days as a function of latitude ($X$ axis).
 }
   \end{figure}
\clearpage
 \begin{figure}
  \centerline{
  \includegraphics[angle=90, width=39pc]
    {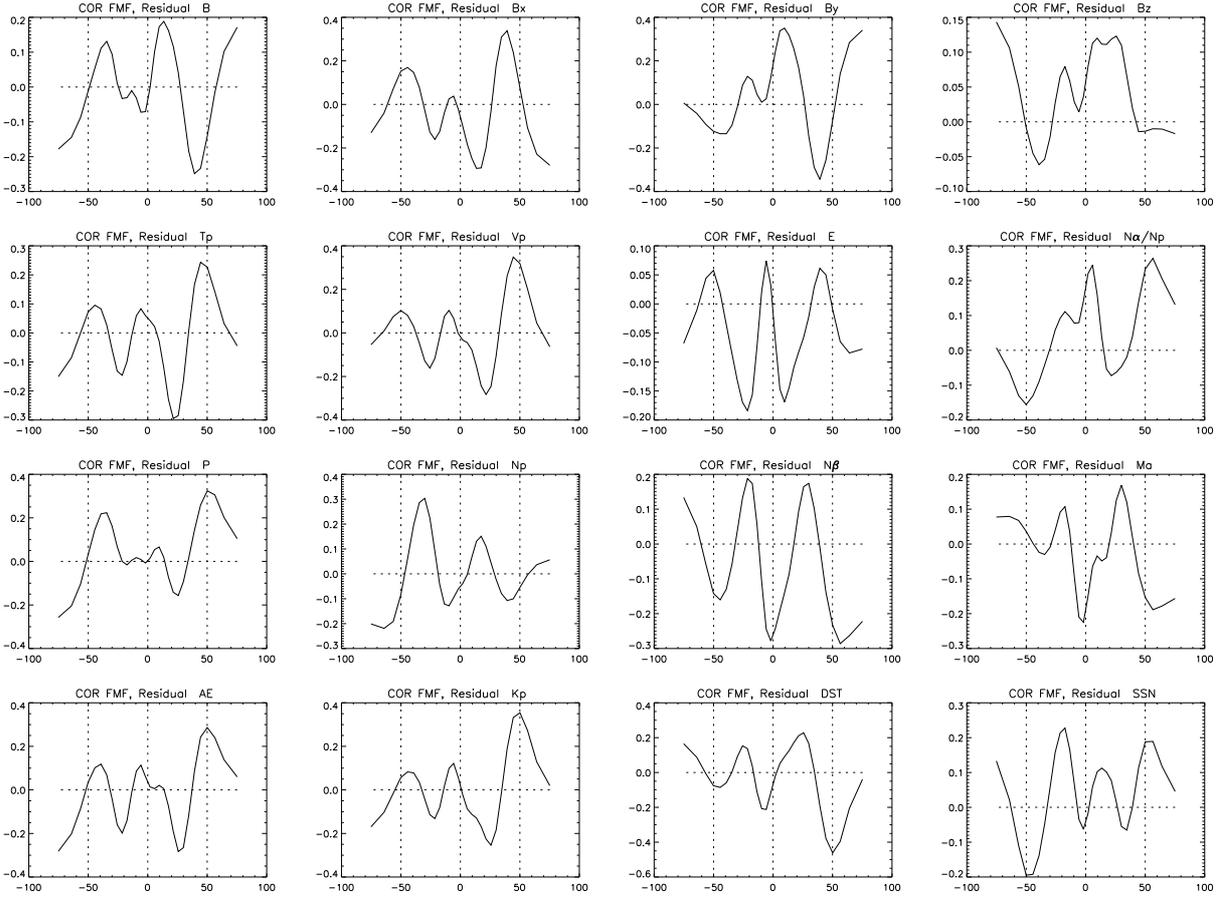}
 }
 \caption{
  Coefficient of correlation between  short term variabilities
  of the photospheric field at different latitudes and
  the interplanetary magnetic field,
  solar wind and geomagnetic parameters with the fixed delay 
  of 4 days as a function of latitude ($X$ axis).
 }
  \end{figure}
\clearpage
 \begin{figure}
  \centerline{
  \includegraphics[angle=90, width=39pc]
    {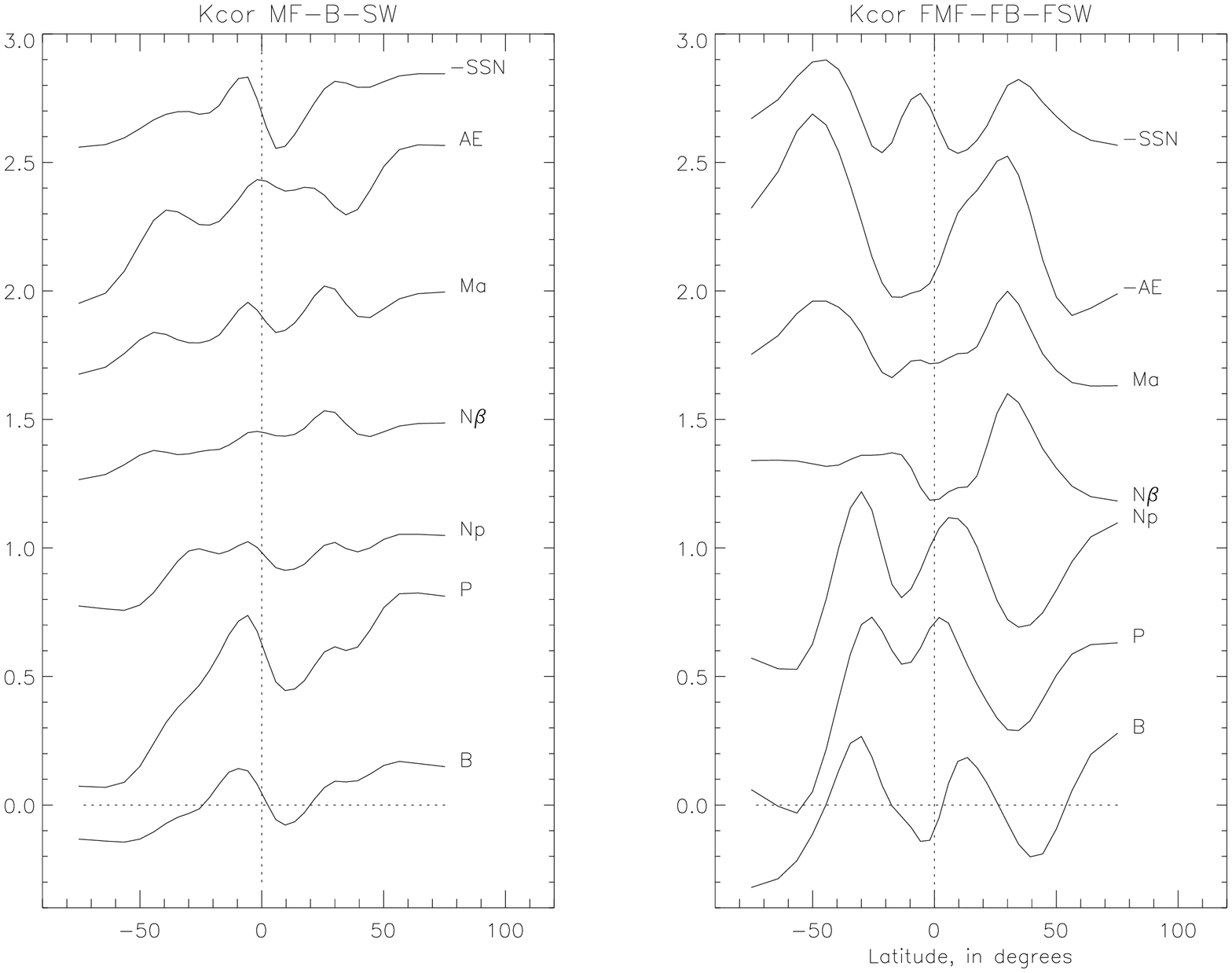}
 }
\caption{
  Coefficients of correlation between 
  the photospheric field and
  the intensity of the interplanetary magnetic field,
  some solar wind and geomagnetic parameters
  (marked on the right end of the corresponding curve)
  for the mean values over 1 year (on the left plot)
  and for the short   term variable  part of them (on the right plot)
  at different latitudes with the fixed delay of 4 days.
 }
  \end{figure}
\clearpage
 \begin{figure}
  \centerline{
  \includegraphics[angle=90, width=39pc]
    {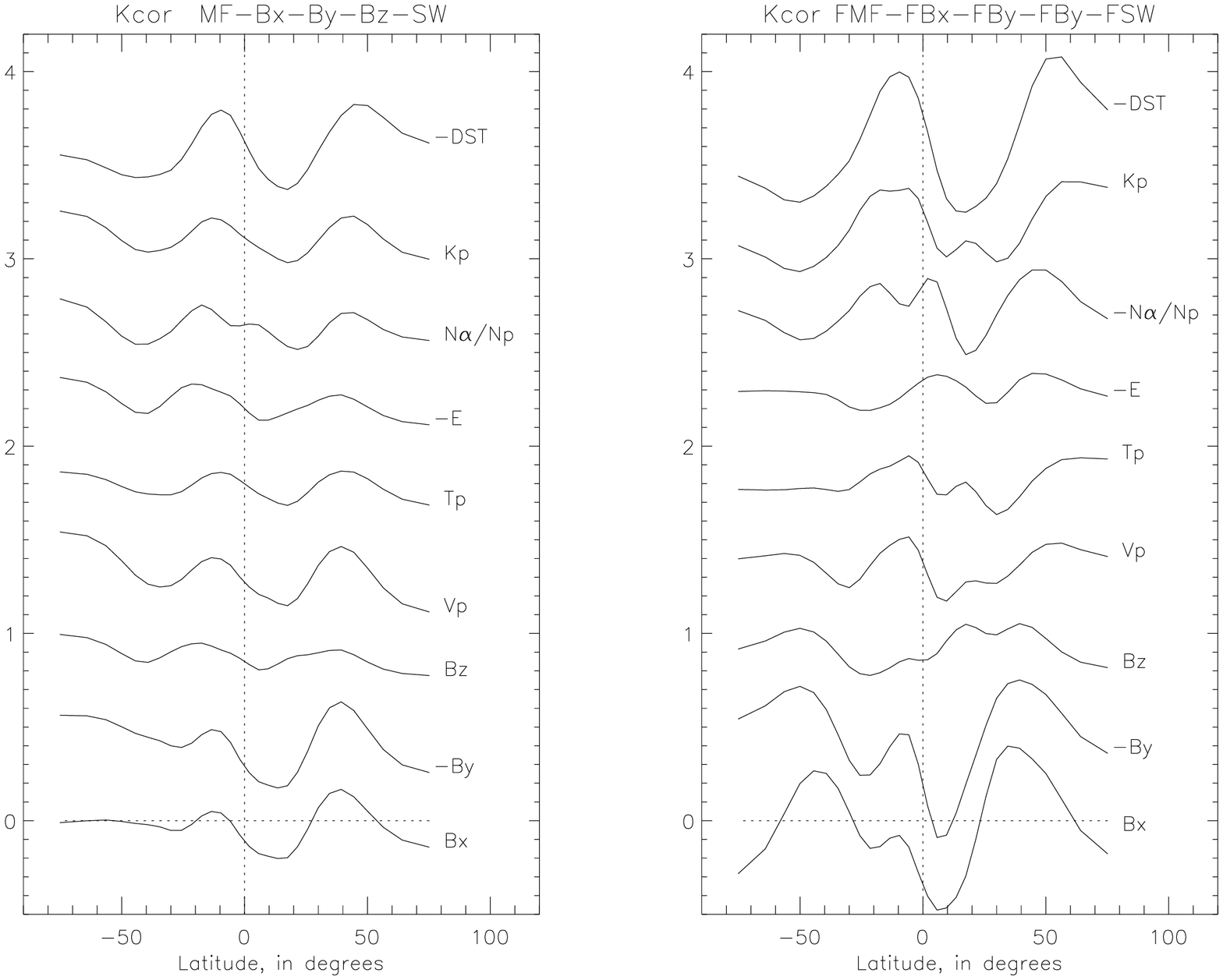}
}  
\caption{
  Coefficients of correlation between 
  the photospheric field and
  the $B_x$, $B_y$ and  $B_z$ components
  of the interplanetary magnetic field,
  some solar wind and geomagnetic parameters
  (marked on the right end of the corresponding curve)
  for the mean values over 1 year (on the left plot)
  and for the short   term variable  part of them (on the right plot)
  at different latitudes with the fixed delay of 4 days.
 }
  \end{figure}


\begin{thebibliography}{}
  \item
  Axford, W.L. and McKenzie J.F. (1997),
  The solar wind,
  in Cosmic winds and the heliosphere,
   ed. by J.R. Jokipii, C.P. Sonett \& M.S. Giampapa,
    31.  

\item
Clua de Gonzalez, W. D. Gonzalez, S. L. G. Dutra, B. T. Tsurutani (1993),
Periodic Variation in the Geomagnetic Activity: A Study Based on the Ap Index,
J. Geophys. Res., 98, 9215.

  \item
  Durney, J.W.   (1961),
  Interplanetary magnetic field and the auroral zones,
   Phys. Rev. Lett., 6, 47.  

\item
Fraser-Smith, A. C. (1973), 
Solar cycle control in the 27-day variation of geomagnetic activity,
J. Geophys. Res., 78, 5825.

\item
   Gavryuseva, E. (2005), Latitudinal streams of solar magnetic field,
    Proc. of 11 Int. Scientific Conf. Solar-Terrestrial Influences,  Nov. 2005,
    BAS, 229-233. 

  \item
  Gavryuseva, E. (2006),
  Topology and dynamics of the  magnetic field
  of the Sun,
   News of the Academy of Science, IzvRAN, ser. Physics,
   70, No. 1, 102.  

\item
  Gavryuseva, E. (2006a),
Latitudinal Structure of the Photospheric Magnetic Field through solar cycles
Solar Activity and its Magnetic Origin, Proc. of the 233rd Symposium of the
IAU, Cairo, Egypt, March 31 - April 4, 2006,
Ed. V. Bothmer; A. A. Hady. Cambridge:
       Cambridge University Press, 124.  

\item
  Gavryuseva, E. (2006b),
Basic topology and dynamics of magnetic field leading activity the Sun
Solar Activity and its Magnetic Origin, Proc. of the 233rd Symposium of the
IAU, Cairo, Egypt, March 31 - April 4, 2006,
Ed. V. Bothmer; A. A. Hady. Cambridge:
       Cambridge University Press, 67.  

\item
  Gavryuseva, E. (2006c),
Variability of the differential rotation of the photospheric magnetic field through solar cycles
Solar Activity and its Magnetic Origin, Proc. of the 233rd Symposium of the
IAU, Cairo, Egypt, March 31 - April 4, 2006,
Ed. V. Bothmer; A. A. Hady. Cambridge:
       Cambridge University Press, 65.  

\item
  Gavryuseva, E. (2006d),
North-South asymmetry of the photospheric magnetic field
Solar Activity and its Magnetic Origin, Proc. of the 233rd Symposium of the
IAU, Cairo, Egypt, March 31 - April 4, 2006,
Ed. V. Bothmer; A. A. Hady. Cambridge:
       Cambridge University Press, 63.  

\item
  Gavryuseva, E. (2006e),
Longitudinal structure of the photospheric magnetic field
Solar Activity and its Magnetic Origin, Proc. of the 233rd Symposium of the
IAU, Cairo, Egypt, March 31 - April 4, 2006,
Ed. V. Bothmer; A. A. Hady. Cambridge:
       Cambridge University Press, 61.  

\item
  Gavryuseva, E. (2006f),
Relationships between photospheric magnetic field, solar wind and geomagnetic perturbations
over last 30 years
Solar Activity and its Magnetic Origin, Proc. of the 233rd Symposium of the
IAU, Cairo, Egypt, March 31 - April 4, 2006,
Ed. V. Bothmer; A. A. Hady. Cambridge:
       Cambridge University Press, 291.  

\item
  Gavryuseva, E. (2008a),
  In search of the origin of the latitudinal structure of the photospheric magnetic field,
  ASP Conf. Ser., v. 383,
  Proc. of "Subsurface and atmospheric influence on solar activity", held at
  NSO, Sacramento Peak, Sunspot, New Mexico, USA
  16-20 April 2007,
  Ed. R. Howe, R. W. Komm, K. S. Balasubramaniam \& G. J. D. Petrie,
  99.  

\item
  Gavryuseva, E. (2008b),
  Longitudinal structure originated in the tachocline zone of the Sun,
  ASP Conf. Ser., v. 383,
  Proc. of "Subsurface and atmospheric influence on solar activity", held at
  NSO, Sacramento Peak, Sunspot, New Mexico, USA
  16-20 April 2007,
  Ed. R. Howe, R. W. Komm, K. S. Balasubramaniam \& G. J. D. Petrie,
  381.  

\item
   Gavryuseva, E. (2018),
     Latitudinal structure and dynamic of the photospheric magnetic field,
    arXiv:1802.02450.   
\item
    Relations between variability of solar and interplanetary characteristics
\item
 To the connection between intensity of the solar and geomagnetic perturbations,
   arXiv:1802.NNNN(N) 
\item
   Gavryuseva, E. (2018),
     Latitudinal structure and dynamic of the photospheric magnetic field,
    arXiv:1802.02450
\item
   Gavryuseva, E. (2018b),
   Longitudinal structure of the photospheric magnetic field in Carrington system,
   arXiv:1802.02461   

\item
Gavryuseva, E.; Godoli, G.  (2006),
Structure and rotation of the large scale solar magnetic field
 observed at the Wilcox Solar Observatory
Physics and Chemistry of the Earth, v. 31, issue 1-3,  68.  

\item
  Gavryuseva, E., Kroussanova, N. (2003),
Topology and dynamics of the Sun's magnetic field
SOLAR WIND TEN: Proceedings of the Tenth International Solar Wind Conference,
AIP Conference Proceedings, v. 679,  242.  

\item
    Gavryuseva, E., and  V. Gavryusev (1994),
      Time variations of the $^{37}Ar$ production rate
      in chlorine solar neutrino experiment,
 Astron. Astrophys, 283, 978.  
  
\item
   Gavryuseva, E., and V. Gavryusev (2000),
   Solar variability and its prediction,
   Long and short term variability
   in Sun's history and global change,
   ed. W.Schroder, Science Edition, Bremen, Germany, p.89.  

 \item
   Gavryuseva, E., and N. Kroussanova,  (2003),
   Topology and dynamic of solar magnetic field,
   Proc. of the Tenth International Solar Wind Conference, 
    AIP Conf. Proc., v. 679, 242.  

\item
     Gavryusev, V., E. Gavryuseva, Ph. Delache, and F. Laclare (1994),
     Periodicities in solar radius measurements,
     Astron. Astrophys, 286, 305.  

\item
   Gavryuseva, E., V. Gavryusev, and M.P. Di Mauro (2000),
   Internal rotation of the Sun as inferred from GONG observations,
   Astronomy Lett., 26, N 4, 261.
 \item
    Gavryuseva, E., \& G. Godoli (2006),
    Structure and rotation of the large scale solar
    magnetic field observed at the Wilcox Solar Observatory,
     Physics and Chemistry of the Earth, Elsevier, 
     31, 68. 
 \item
   Gonzalez, W.D., J.A. Joselyn, Y. Kamide, H.W. Krorhl, G. Rostoker,
   B.T. Tsurutani and V.M. Vasyliunas (1994),
   What is a geomagnetic storm?,
   J. Geophys. Res., 99, 5771.  

 \item
   Gonzalez, W.D., B.T. Tsurutani and A.L. Clua de Gonzalez (1999),
   Interplanetary origin of geomagnetic storms,
   Space Science Rev., 88,
   529.  

 \item
   Kane, R.P. (2005a),
   Difference in the quasi-biennial oscillation 
   and quasi-triennial oscillation characteristics of the solar,
   interplanetary, and terrestrial parameters,
   J. Geophys. Res., 110, A01108.  

 \item
   Kane, R.P. (2005b),
   Short-term periodicities in solar indices,
   Solar Phys., 227, 155.  

 \item
   Li, Y., Luhmann, J. G., Arge, C. N., Ulrich, R., How do solar magnetic fields 
   influence the long term changes of some geomagnetic indexes?, American
   Geophysical  Union, Spring Meeting 2001, abstract SH52A-02, 2001.  

 \item
   Pizzo, V. J., A three-dimensional model of corotating streams in the solar
   wind. III Magnetohydrodynamic streams, J. Geophys. Res., 87, 4374, 1982.  

 \item
   Wang, Y.-M., J. Lean, and N. R. Sheeley, The long-term variation of the 
   Sun's open magnetic flux, Geophys. Res. Lett., 27, 505, 2000.  

 \item
   Low, B.C.  (1996),
   Solar activity and the corona,
   Solar Phys., 167, 217.  

 \item
   Luhmann, J. G., Li, Y., Arge, C. N., Gazis, P. R., Ulrich, R., Solar cycle
   changes in coronal holes and space weather cycles, J. Geophys. Res., 107(A8),
   1154, pp. SMP 3-1, 2002.  

 \item
   Makarov, V. I., Tlatov, A. G., Callebaut, D. K., Obridko, V. N., Increase of 
   the Magnetic Flux From Polar Zones of the sun in the Last 120 Years, Solar
   Physics, v. 206, Issue 2, p. 383-399 (2002).  

 \item
   Parker, E.N., (1997),
   Mass ejection and a brief history of the solar wind concept,
   in Cosmic winds and the heliosphere,
   ed. by J.R. Jokipii, C.P. Sonett and M.S. Giampapa,
   p.3.  

\item
   Rivin, Yu.R. (1989),
   Cycles of the Earth and of the Sun,
   Nauka, IZMIRAN, p.36.  

 \item
   Smith, E.J., (1997),
   Solar wind magnetic field,
   in Cosmic winds and the heliosphere,
   ed. by J.R. Jokipii, C.P. Sonett and M.S. Giampapa, p.425.  

 \item
    Stamper E.J., (1997),
    Solar wind magnetic field,
    in Cosmic winds and the heliosphere,
    ed. by J.R. Jokipii, C.P. Sonett and M.S. Giampapa, p.425.  

 \item
    Stamper, R., Lockwood, Wild, M.N., Clark, T.D.G., (1999),
    Solar causes of the long-term increase of the geomagnetic
    activity, J. Geophys. Res., 104, Issue A12, pp.28,325.

\item
Pizzo, V. J. (1982),
    A three-dimensional model of corotating streams in the solar wind.
    III Magnetohydrodynamic streams,
    J. Geophys. Res., 87, 4374.

\item
    Scherrer, P.H., J.M. Wilcox, L.Svalgaard,
    T.L. Duvall, Ph.H. Dittmer and E.K. Gustafson (1977),
    The mean magnetic field of the Sun: observations at Stanford.
    Solar Phys., 54, 353.  
%

\item
Wang, Y.-M., J. Lean, and N. R. Sheeley (2000),
  The long-term variation of the Sun's open magnetic flux,
   Geophys. Res. Lett., 27, 505.
 \end{thebibliography}
\end{document}